\documentclass[prb,preprint,showpacs,amsmath,amssymb]{revtex4}

\usepackage{graphicx}
\usepackage{dcolumn}
\usepackage{bm}

\begin{document}

\title{A perturbative approach to $J$ mixing in $f$-electron systems:
Application to actinide dioxides}

\author{N. Magnani}
\author{P. Santini}
\author{G. Amoretti}

\affiliation{I.N.F.M. and Dipartimento di Fisica, Universit\`a di
Parma, Parco Area delle Scienze 7/A, I-43100 Parma, Italy}

\author{R. Caciuffo}
\affiliation{I.N.F.M. and Dipartimento di Fisica ed Ingegneria dei Materiali e
del Territorio, Universit\`a Politecnica delle Marche, Via Brecce Bianche,
I-60131 Ancona, Italy}


\begin{abstract}

We present a perturbative model for crystal-field calculations,
which keeps into account the possible mixing of states labelled by
different quantum number $J$. Analytical $J$-mixing results are
obtained for a Hamiltonian of cubic symmetry and used to interpret
published experimental data for actinide dioxides. A unified
picture for all the considered compounds is proposed by taking
into account the scaling properties of the crystal-field
potential.

\end{abstract}


\maketitle

\section{\label{sec:intro}Introduction}

Crystal-field (CF) theory\cite{newmanbook} is one of the most
powerful theoretical methods to deal with the magnetic properties
of rare-earth (RE) and actinide (An) ions, and Stevens' operator
equivalents formalism is still the most commonly used to analyze
experimental data due to its simplicity. Unfortunately, this
approach concentrates only on the CF splittings within the
lowest-lying $^{2S+1}L_{J}$ multiplet of the considered ion,
completely neglecting the contributions of excited multiplets
(``$J$ mixing''). Although the task of diagonalizing the large
matrices related to the full $f^{n}$ configuration, including
different $J$ multiplets, is relatively easy to perform
numerically by means of today's computers, Stevens' approach often
makes it possible to obtain analytical expressions for physical
quantities of interest for systems of sufficiently high symmetry,
thus leading to a deeper insight on the physics of several
compounds.

In the present paper, we discuss a perturbative approach which
retains the validity of Stevens' formalism while correctly taking
into account $J$-mixing effects. This method, which has led to
interesting results for transition-metal (TM) based molecular
clusters\cite{liviotti,prlfe8} and ferromagnetic exchange-driven
RE-TM intermetallic compounds\cite{magnani} is now applied to
evaluate the intramultiplet CF splittings in light An and RE ions.
We exploit the method to analyze the CF of actinide dioxides.
These large-gap semiconductors are among the most studied actinide
compounds. Although $f$-electrons are well localized, the
complexity of the magnetic Hamiltonian, which includes CF and
magnetoelastic single-ion interactions, phonon-transmitted
quadrupolar interactions, and multipolar superexchange couplings
between neighboring ions, leads to a number of interesting and
unusual physical phenomena. Among them, we mention the proposed
octupolar phase transition in NpO$_2$,\cite{oct1,oct2} the
observed CF-phonon bound states in NpO$_2$,\cite{bound} and the
peculiar static and dynamic phenomena produced by magnetoelastic
interactions in UO$_2$,\cite{sasaki,cowley,uo299} some of which
are not yet fully understood.

The CF potential is the fundamental building block of any
theoretical model of the properties of dioxides, since this
influences the single-ion behavior to a large extent. In
particular, it determines which degrees of freedom of the $f$
shell are left unquenched and the size of the corresponding
multipole moments, which account for the low-$T$ physical
properties. Most of the published theoretical approaches are based
on the above-mentioned Stevens' treatment of the CF, which
includes only the lowest Russell-Saunders or Intermediate-Coupling
multiplet of the ion. If one takes as starting point the CF of
UO$_2$, on which very detailed information is available by
inelastic neutron scattering (INS) experiments,\cite{amoretti}
then scaling the CF of UO$_2$ within the Stevens' framework (to
take into account the different ionic radii) provides a good CF
model for NpO$_2$. However, the same scheme applied to PuO$_2$ is
only qualitatively satisfactory, since it reproduces the correct
level sequence but it underestimates the observed energy
splitting.

Moreover, this approach is not internally consistent because the
so-obtained CF parameters yield different results when additional
ionic multiplets are included in the calculation. On the other
hand, the increased complexity of $J$ mixing calculations makes
particularly hard to find CF parameter sets working consistently
over the various compounds. Indeed, different sets have been
proposed so far for dioxides within $J$ mixing calculations. In
particular, in NpO$_2$ two distinct and equally good sets of
parameters had been obtained.\cite{amorettinpo2}

By our perturbative $J$ mixing approach, we have been able to
obtain a unique set which works well over all the considered
compounds.

\section{\label{sec:model}The perturbative $J$-mixing model}

Following Ref.~\onlinecite{slichter}, the total free-ion and
crystal-field Hamiltonian $H=H_{FI}+H_{CF}$, with
\begin{equation}
H_{CF}=\sum_{k,q}B_{k}^{q}C_{q}^{(k)} \label{hcfgeneral}
\end{equation}
can be rewritten in the form
\begin{equation}
H=H_{0}+H_{1}+H_{2},
\end{equation}
where in the present case $H_{0}$ coincides with $H_{FI}$ and
$H_{1}$ and $H_{2}$ are chosen so that the former has nonzero
matrix elements only between states belonging the the same
$^{2S+1}L_{J}$ multiplet. It is possible to define a Hermitian
operator $\Omega$ such that the matrix element of the transformed
Hamiltonian $H^{\prime}=e^{-i\Omega}He^{i\Omega}$ are very small
in the off-diagonal blocks, thus restoring the possibility to use
an isolated-multiplet approach. In this framework,\cite{slichter}
\begin{equation}
\left<\alpha J M\left|H^{\prime}\right|\alpha J M^{\prime}\right>=
E_{0 \alpha J}\delta_{MM^{\prime}}+
\left<\alpha J M\left|H_{1}\right|\alpha J M^{\prime}\right>
\label{daslichter}
\end{equation}
\[
-
\sum_{\alpha^{\prime \prime} J^{\prime \prime} M^{\prime \prime}}
\frac{\left<\alpha J M\left|H_{2}\right|\alpha^{\prime \prime} J^{\prime \prime} M^{\prime \prime}\right>
\left<\alpha^{\prime \prime} J^{\prime \prime} M^{\prime \prime}\left|H_{2}\right|\alpha J M^{\prime}\right>}
{E_{0 \alpha^{\prime \prime} J^{\prime \prime}}-E_{0 \alpha J}},
\]
where $\alpha, J$ label free ion manifolds and $E_{0 \alpha J}$
are the eigenvalues of $H_0$. For clarity, in the following we
will label the states as in the Russell-Saunders scheme, where
$\alpha$ coincides with $(L,S)$ and any additional quantum number
necessary to identify the terms; yet the actual calculation will
include Intermediate Coupling corrections to the eigenfunctions.

Once we limit our calculations to the ground $J$ multiplet only,
the first term on the right-hand side of the above equation
represents a uniform energy shift of the whole multiplet, while
the second term is the usual ground-multiplet CF Hamiltonian $H^{(J)}$.
The effect of $J$-mixing is accounted for by the third term, which
will be considered as an extra contribution to the ground-multiplet
Hamiltonian and labelled $H^{(J)}_{mix}$ (we maintain the
redundant superscript $(J)$ notation in order to emphasize that
the newly obtained $J$-mixing Hamiltonian {\it also} acts on the
ground multiplet {\it only}).

In the case of light actinides, the most important $J$ mixing contribution comes from
the two lowest $J$-multiplets, i.e. $^{2S+1}L_{J}$ and $^{2S+1}L_{J+1}$.
Therefore, for the sake of simplicity we restrict our analysis to the case of these two multiplets,
separated by an energy gap $\Delta$ by the spin-orbit
interaction. From Eq.~\ref{daslichter}, $H^{(J)}_{mix}$ can be written as
\[
\left< J M \left| H^{(J)}_{mix} \right| J M^{\prime} \right> = -
\sum_{M^{\prime \prime}} \sum_{k,q}\sum_{k^{\prime},q^{\prime}}
B_{k}^{q} B_{k{\prime}}^{q{\prime}}
\]
\begin{equation}
\times \frac{
\left< J M \left| C_{q}^{(k)} \right| J+1 M^{\prime \prime} \right>
\left< J+1 M^{\prime \prime} \left| C_{q^{\prime}}^{(k^{\prime})} \right| J M^{\prime} \right>}
{\Delta}.
\label{mix1}
\end{equation}
The Wigner-Eckart theorem, in the form
\begin{equation}
\left\langle J_{1}M_{1}\left| C_{q}^{(k)}\right| J_{2}M_{2}\right\rangle
=\left( -1\right) ^{J_{1}-M_{1}}\left\langle J_{1}\left\| C^{(k)}\right\| J_{2}
\right\rangle \left(
\begin{array}{ccc}
J_{1} & k & J_{2} \\
-M_{1} & q & M_{2}
\end{array}
\right)
\label{wigeck}
\end{equation}
allows us to get rid of the sum over $M^{\prime \prime}$ in
Eq.~\ref{mix1}, since the $3j$ symbol in Eq.~\ref{wigeck} equals
zero if $M_{2} \neq M_{1} - q$. The products of $3j$ symbols can
be rewritten as linear combinations of matrix elements of Stevens
operators $O_{k}^{q}$,\cite{liviotti,magnani} so that
\begin{equation}
H^{(J)}_{mix} = \sum_{k,q,k^{\prime},q^{\prime}}
\frac{B_{k}^{q} B_{k^{\prime}}^{q^{\prime}}}{\Delta}M_{k+k^{\prime}}^{q+q{\prime}}
\end{equation}
where the ``mixing operators'' $M_{n}^{m}$ are\cite{liviotti}
\begin{equation}
M_{n}^{m}=\sum_{p=0}^{n} c_{n,p}^{(m)} O_{p}^{m}
\end{equation}
with conveniently defined $c_{n,p}^{(m)}$ coefficients.

Let us consider the simple but important case of cubic symmetry,
for which the CF Hamiltonian (\ref{hcfgeneral}) has the form
\begin{equation}
H_{CF}=B_{4}^{0}\left[ C_{0}^{(4)}+\sqrt{\frac{5}{14}}\left(
C_{4}^{(4)}+C_{-4}^{(4)}\right) \right]
+B_{6}^{0}\left[ C_{0}^{(6)}-\sqrt{%
\frac{7}{2}}\left( C_{4}^{(6)}+C_{-4}^{(6)}\right) \right] .
\label{hcubic}
\end{equation}
Restricting the calculations within the ground multiplet only and using the
Stevens' operator equivalents formalism, Eq.~(\ref{hcubic}) becomes
\begin{equation}
H_{CF}^{(J)}=\frac{B_{4}^{0}}{8}\beta \left( O_{4}^{0}+5O_{4}^{4}\right) +%
\frac{B_{6}^{0}}{16}\gamma \left( O_{6}^{0}-21O_{6}^{4}\right) ,
\label{hcfj}
\end{equation}
$\beta$ and $\gamma$ being the fourth- and sixth-order Stevens factors.
To introduce another common notation,\cite{newmanbook} we define
\begin{equation}
V_{4}=\frac{B_{4}^{0}}{8}=A_{4}\left< r^{4} \right> ;
V_{6}=\frac{B_{6}^{0}}{16}=A_{6}\left< r^{6} \right> ,
\label{notation}
\end{equation}
where $\left< r^{n} \right>$ are the expectation values of the
$r^{n}$ operator over the appropriate $f$-electron wavefunction.
It is found that $H^{(J)}_{mix}$ maintains the cubic symmetry, and
has the form
\begin{equation}
H_{mix}^{(J)}=\nu _{4}\left( O_{4}^{0}+5O_{4}^{4}\right)
+\nu_{6}\left( O_{6}^{0}-21O_{6}^{4}\right)
+\nu _{8}\left(
O_{8}^{0}+28O_{8}^{4}+65O_{8}^{8}\right) + \ldots ,
\label{hmixcub}
\end{equation}
where we did not explicitly write the terms containing operators
of rank higher than 8 since $O_{k}^{q} = 0$ for $k > 2J$, and
$J=9/2$ is the maximum possible value for the ground state of
light lanthanides and actinides.\cite{notaopstevens} The
coefficients appearing in Eq.~(\ref{hmixcub}) are dependent on $J$
and can be written as
\[
\nu _{k}=\frac{1}{\Delta} \left[
\nu _{k}^{(4,4)}\left( V_{4}\left\langle J\left\| C^{(4)}\right\|
J+1\right\rangle \right) ^{2}
+\nu _{k}^{(6,6)}\left( V_{6}\left\langle
J\left\| C^{(6)}\right\| J+1\right\rangle \right) ^{2}
 \right.
\]
\begin{equation}
\left. +\nu_{k}^{(4,6)} \left( V_{4}\left\langle J\left\|
C^{(4)}\right\| J+1\right\rangle V_{6}\left\langle J\left\|
C^{(6)}\right\| J+1\right\rangle \right) \right] ,
\label{coefficient}
\end{equation}
and a list of $\nu _{k}^{(m,n)}$ for the ground multiplets of
$f^{n}$ configurations with $1 \leq n \leq 5$ is given in
Table~\ref{tab:table1}. The reduced matrix elements $\left\langle
J\left\| C^{(k)}\right\| J+1\right\rangle$ are calculated by using
the Intermediate Coupling free-ion wavefunctions.

Ions with 6 $f$ electrons have a $J=0$ ground singlet, so that no
intramultiplet energy splitting can exist and $J$-mixing effects
are evident only in the wavefunction composition; this case is
then impossible to study by the present approach. As for ions with
half-filled $f$ shell such as Gd$^{3+}$, Cm$^{3+}$, and Bk$^{4+}$,
$J$ mixing is generally negligible and it hardly affects any
physical property. Finally, the perturbative $J$-mixing approach
is in principle suitable to study heavy $f$-electron ions ($8 \leq
n \leq 13$), taking into account that the ground $J$ multiplet is
mixed with $J-1$ states, instead of $J+1$. However, in this case
the advantage of using the perturbative model with respect to the
numerical diagonalization of $H$ over a complete $f^{n}$ basis
could be significantly reduced, because: {\it i}) the use of
Stevens operators of rank 10 (12) becomes necessary for $J \geq 5$
(6); {\it ii}) it is not always possible to diagonalize
$H^{(J)}_{CF}+H^{(J)}_{mix}$ analytically, even in cubic symmetry.
In any case, the expected $J$-mixing strength is much smaller for
heavy than for light elements.

\section{\label{sec:f1}A particular case: the $f^{1}$ configuration}

In order to show in detail how the present model can be applied to the
study of rare-earth and actinide compounds, let us start from the simplest
possible configuration: an ion with a single $f$ electron.
The only interaction present in the free-ion Hamiltonian is the spin-orbit
coupling,
\begin{equation}
H_{FI}=\Lambda {\bf L\cdot S}.
\end{equation}
The $f^{1}$
spectra is composed of two multiplets only, $^{2}F_{5/2}$ (ground state) and $^{2}F_{7/2}$;
adding a crystal field of cubic symmetry, the $14\times 14$ matrix
representing the $^{2}F$ term is made up of two $4\times 4$ and two $3 \times 3$
diagonal blocks. The complete Hamiltonian $H$ can then be
analytically diagonalized, and the resulting energy gap between the
$\Gamma_{7}$ doublet and a $\Gamma_{8}$ quartet composing the
ground multiplet is
\[
E_{\Gamma_{7}}-E_{\Gamma_{8}} = \frac{1}{1716} \bigg[
3744V_{4}-4480V_{6}
\]
\[
+\sqrt{25600\left( 13V_{4}-84V_{6}\right)^{2}
-137280\left( 13V_{4}-84V_{6}\right)\Lambda+
9018009\Lambda^{2}}
\]
\begin{equation}
-\sqrt{16384\left(13V_{4}+70V_{6}\right)^{2}-
219648\left(13V_{4}+70V_{6}\right)\Lambda+9018009
\Lambda^{2}}~\bigg] . \label{f1compl}
\end{equation}
Although the perturbative approach is not particularly useful here
since an exact analytical solution can be obtained, let us study
this simple case in order to clarify the details of the process
and to understand its limits of validity. No intermediate coupling
occurs, and $\beta=2/315$, $\gamma=0$, $\left< J \left\| C^{4}
\right\| J+1 \right> = 2 \sqrt{10/77}$, and $\left< J \left\|
C^{6} \right\| J+1 \right> = -10 \sqrt{2/143}$; moreover, the
spin-orbit gap can be expressed as $\Delta=\Lambda(J+1)=7\Lambda
/2$, with $\Lambda = 100.5~{\rm meV}$. Diagonalizing
$H_{CF}^{(J)}+H_{mix}^{(J)}$ (see Appendix for details), we find
\begin{equation}
E_{\Gamma_{7}}-E_{\Gamma_{8}} = \frac{16}{7}V_{4} +
\frac{1}{\Lambda}\left(
\frac{20480}{124509}V_{4}^{2}
 -\frac{614400}{77077}V_{4}V_{6}+
\frac{204800}{20449}V_{6}^{2} \right) ,
\label{f1appr}
\end{equation}
which corresponds exactly to the series expansion of
Eq.~(\ref{f1compl}) up to the first order in $\Lambda^{-1}$.

The intramultiplet energy gap calculated above can be directly
measured by means of spectroscopic techniques; for example,
inelastic neutron scattering measurements for PrO$_{2}$ (where
praseodymium ions have valence 4+, therefore presenting a $4f^{1}$
electronic configuration) have shown that
$E_{\Gamma_{7}}-E_{\Gamma_{8}} = 131~\rm{meV}$.\cite{boothroyd}
Figure~\ref{fig:fig1} shows the possible solutions of this
equation in terms of the crystal-field parameters $V_{4}$ and
$V_{6}$, with three different expressions: the full black line
corresponds to the exact diagonalization of the complete $f^{1}$
Hamiltonian [Eq.~(\ref{f1compl})]; the dashed vertical line is
obtained by neglecting $J$ mixing and using Stevens' approximation
$E_{\Gamma_{7}}-E_{\Gamma_{8}} = (16/7)V_{4}$; finally, the
dashed-dotted line takes $J$ mixing into account perturbatively by
the present approach [Eq.~(\ref{f1appr})]. The results are
satisfactory: the agreement between the exact (full) and
approximate (dashed-dotted) curves is qualitatively much better if
compared with Stevens' approximation (dashed line), and the
quantitative contribution of the excited states is correctly
estimated in the small-CF range $\left| V_{k}/\Lambda\right|<1$.
It may be noticed that the three curves of Fig.~\ref{fig:fig1}
almost coincide when $V_{6} \simeq 0$, since in this case the
largest part of the $J$-mixing correction (which is linear in
$V_{6}$) vanishes [this can be verified by observing the relative
magnitude of the different coefficients in Eq.~(\ref{f1appr})].
For the same reason, for small $V_{6}$, the value of $V_{4}$ is
slightly underestimated for $V_{6}<0$ and overestimated for
$V_{6}>0$. As the spin-orbit interaction is stronger for heavier
ions and the gap between the two lowest multiplets grows with $J$,
we expect our model to have a reasonably good performance over the
whole lanthanide and actinide series.

\section{\label{sec:ano2}The cubic phase of actinide dioxides}

In this Section, the perturbative $J$-mixing model outlined so far
will be applied to interpret the intramultiplet crystal-field
splittings observed by inelastic neutron scattering (INS) for
actinide dioxides AnO$_{2}$ (An = U, Np, Pu). The crystal-field
analysis will be performed in terms of the parameters $A_{4}$ and
$A_{6}$ [Eq.~(\ref{notation})] instead of $V_{4}$ and $V_{6}$,
using the values of $\left< r^{n} \right>$ given in
Ref.~\onlinecite{raggimedi}. Although recent density functional
studies have pointed out a certain degree of covalency for the
$An$-O bond in these systems,\cite{wu} this does not prevent one
from using the crystal field theory to analyze experimental data;
it would be quite more difficult than in a ionic compound to
calculate the CF parameters from first principles, but this is not
our aim. We will demonstrate that the present method can be used
for an analytical study of the experimental results over a wide
range of parameters and compositions.

INS spectra for UO$_{2}$ in the paramagnetic phase (above $T_{N} =
30.8 K$) display peaks between 150 ad 185
meV,\cite{kernuo2,amoretti} which have been attributed to
transitions between the $\Gamma_{5}$ ground
state\cite{kernuo2,amoretti,rahman} and excited $\Gamma_3$ and
$\Gamma_4$ states, and no other magnetic transitions were reported
up to 700 meV\cite{amoretti} (it is worth to recall that the
$\Gamma_{5} \rightarrow \Gamma_{1}$ transition is not
dipole-allowed so, if present, it will display a very small
intensity with respect to the other two transitions).

Figure~\ref{fig:fig2} shows the values of $A_{4}$ and $A_{6}$ for
which the possible transitions lie within the experimentally
observed range according to the perturbative model. In the
previous Section, we have studied the quantitative discrepancy
between the model's predictions and the exact results, which was
found to be significant for large CF parameters. Following these
estimates, we have determined a ``safe zone'' (indicated in
Fig.~\ref{fig:fig2} by a dashed ellipse) within which the true set
of CF parameters for UO$_2$ is located with high degree of
confidence.

For the paramagnetic phase of NpO$_{2}$ we have followed Amoretti et
al.,\cite{amorettinpo2} who observed a broad magnetic signal centered at 55 meV in
the INS spectra and attributed this peak to a transition between the two $\Gamma_{8}$
quartets. They gave two possible solutions for the paramagnetic phase, labelled 3
and 4 in Fig.~\ref{fig:fig3}; we show that actually there are infinite possible solutions,
divided in two branches.

PuO$_2$ displays a temperature-independend magnetic susceptibility
below 1000 K,\cite{raphael} so that the CF ground state is
expected to be the $\Gamma_{1}$ singlet. Magnetic-dipole matrix
elements involving this state within the $^{5}I_{4}$ multiplet are
zero except with the $\Gamma_{4}$ triplet. Indeed, only one peak
centered at 123 meV is observed in PuO$_{2}$ INS
spectra.\cite{kernpuo2} According to our perturbative model there
are again infinite possible solutions, plotted in
Fig.~\ref{fig:fig3}, which cover a large area of the
$A_{4}$-vs-$A_{6}$ diagram.

It is clear that, in the case of AnO$_{2}$, the INS data analysis
cannot give unambiguous results if every compound is treated
separately; this can also be inferred from the widely scattered
sets of parameters which are found in the literature (some of
which are listed in Table~\ref{tab:table2} and displayed in
Fig.~\ref{fig:fig3}). On the other hand, if we assume that the
$A_{k}$ parameters are approximately the same for all
isostructural compounds,\cite{newmanbook} an inspection of
Fig.~\ref{fig:fig3} shows that the only area of the diagram where
common solutions for $An =$U, Np, Pu might exist is around the
point labelled 4, which corresponds to one of the two solutions
(the ``strong $J$-mixing'' one) proposed by Amoretti et
al.\cite{amorettinpo2} for NpO$_{2}$ ($A_{4}=-19.6~{\rm
meV}/a_{0}^{4}$ ; $A_{6}=0.666~{\rm meV}/a_{0}^{6}$). In order to
verify this result, we have calculated the INS transition energies
for UO$_{2}$, NpO$_{2}$, PuO$_{2}$, and PrO$_{2}$ with this set of
parameters by numerical diagonalization of the complete $f^{n}$
configuration Hamiltonian (Table~\ref{tab:table3}). Moreover, as a
further test, we have calculated the magnetic susceptibility for
AmO$_{2}$ and CfO$_{2}$ with the same parameters
(Fig.~\ref{fig:fig4}; the measurements can be found in
Ref.~\onlinecite{karraker} and Ref.~\onlinecite{moore}
respectively).

In all the cases examined so far, the comparison with experimental
results is quite good, considering that
a $100\%$ exact scaling of the CF potential is not expected to hold.
Therefore, our results point towards a coherent unified picture
for the CF potential in actinide dioxides.

The splitting predicted
for PrO$_2$ seems less satisfactory as the measured value of the
gap $E(\Gamma _{7}) -E(\Gamma _{8}) = 131$ meV is quite larger
than the value calculated with the solution we propose. However,
the PrO$_2$ case is complicated by magnetoelastic interactions
that are known to affect heavily the physics of this compound by
increasing the bare value of the CF gap.\cite{boothroyd} Hence, a
value of this bare gap of the order of 80 meV is fully realistic.

One feature which cannot be accounted for by the CF models
proposed so far is the temperature-independence of the magnetic
susceptibility of PuO$_{2}$ above 600 K. Indeed, it is obvious
that if there is a magnetic gap of 123 meV as observed by neutron
scattering, when this level becomes thermally populated it will
contribute to the susceptibility, no matter the mechanism which
generates the splitting (unless this contribution is accidentally
canceled by other contributions). Indeed, even a completely
different approach based on Density-Functional-Theory calculations
cannot remove the discrepancy between magnetic susceptibility and
neutron scattering experiments.\cite{mamo} Using the diagram in
Fig.~\ref{fig:fig3} as a guide we have been able to find a
solution for PuO$_{2}$ ($A_{4}=-26.7~{\rm meV}/a_{0}^{4}$,
$V_{6}=1.68~{\rm meV}/a_{0}^{6}$) which leads to
$E_{\Gamma_{4}}-E_{\Gamma_{1}}=134~{\rm meV}$ and a flat $\chi
(T)$ curve below 1000 K. For this set the Curie contribution of
the excited triplet and the off-diagonal Van Vleck contribution of
the $\Gamma_{1}-\Gamma_{4}$ pair accidentally combine into an
almost $T$-independent susceptibility. However, this solution is
quite unstable even for small variation of the CF parameters;
moreover, it does not give good results if applied to UO$_{2}$ and
NpO$_{2}$.

\section{Conclusions}

We have developed a perturbative approach to $J$ mixing which
allows the Stevens' formalism to be recovered by replacing the
original CF hamiltonian with an effective one operating within the
same $J$ multiplet. We have applied the method to the study of the
CF in actinide dioxides. These compounds have extremely
interesting physical properties, which are determined by the CF to
a large extent. It is therefore important to reach a precise
understanding of the CF, which is the basic building block of all
theoretical efforts devoted to understand these properties. Most
studies of the CF potential rely on the Stevens' approach, which
yields a solution working fairly consistently over several
compounds. Yet, this solution does not work satisfactorily anymore
when $J$ mixing is included, and the practical impossibility to
perform exact $J$ mixing calculations in wide ranges of the
parameter space has prevented so far the identification of a
better alternative. By hugely decreasing the numerical effort in
favor of analytical calculations, our method has allowed us to
find a $J$ mixing solution ($A_{4}=-19.6~{\rm meV}/a_{0}^{4}$ ,
$A_{6}=0.666~{\rm meV}/a_{0}^{6}$) which works consistently over
all the considered compounds.

Although in the present paper we have only studied cubic systems
and included the two lowest multiplets only (which in dioxides
account for almost 100\% of $J$ mixing), the method can be used
for any point symmetry, at the price of additional terms in the
effective Hamiltonian; also, as many multiplets as necessary can
be included in the calculation, leading to additional
contributions to the $\nu_k$ coefficients of
Eq.~(\ref{coefficient}). Even in this case our method, by allowing
to perform a quantitative analysis of experimental data by means
of a Stevens-like Hamiltonian containing higher-rank operators, is
much more efficient than fully numerical diagonalization. In
particular, wide ranges of the parameter space can be easily
investigated, allowing to produce quite easily diagrams such as
Figs.~\ref{fig:fig2} and~\ref{fig:fig3} which would otherwise be
very hard (if not impossible, in some cases) to obtain.

\appendix*

\section{\label{appcubic}Analytical diagonalization of a cubic crystal-field Hamiltonian}

This Appendix will be devoted to illustrate the results of analytical diagonalization
of the cubic crystal-field Hamiltonian
\begin{equation}
H_{cubic}^{(J)}=\frac{b _{4}}{8}\left( O_{4}^{0}+5O_{4}^{4}\right) +\frac{b
_{6}}{16}\left( O_{6}^{0}-21O_{6}^{4}\right)
+\frac{b _{8}}{128}\left(
O_{8}^{0}+28O_{8}^{4}+65O_{8}^{8}\right) ,
\label{hcfcub}
\end{equation}
which corresponds to $H^{(J)}_{CF}+H^{(J)}_{mix}$ discussed in this paper
[Eqs.~(\ref{hcfj}) and (\ref{hmixcub})] with $b_{4} = 8V_{4}\beta + \nu_{4}$;
$b_{6} = 16V_{6}\gamma + \nu_{6}$; $b_{8} = \nu_{8}$.

For $J=5/2$ (Ce$^{3+}$, Pr$^{4+}$, Sm$^{3+}$, Am$^{4+}$, Pu$^{3+} \ldots$)
$b_{6}$ and $b_{8}$ have no effects,
and the multiplet splits in a $\Gamma_{7}$ doublet and a $\Gamma_{8}$ quartet
separated by $E_{\Gamma_{8}}-E_{\Gamma_{7}} = 45 b_{4}$.

A $J=4$ ground multiplet (Pr$^{3+}$, U$^{4+}$, Np$^{3+}$, Pu$^{4+} \ldots$) is splitted
into a singlet ($\Gamma_{1}$), a doublet ($\Gamma_{3}$) and two triplets ($\Gamma_{4}$
and $\Gamma_{5}$) by a cubic crystal field. The energy separations between these
levels are:
\begin{equation}
E_{\Gamma_{3}}-E_{\Gamma_{1}} = 180(-b_{4}+63b_{6}-21b_{8}) ;
\end{equation}
\begin{equation}
E_{\Gamma_{4}}-E_{\Gamma_{1}} = 105(-b_{4}+63b_{6}-216b_{8}) ;
\end{equation}
\begin{equation}
E_{\Gamma_{5}}-E_{\Gamma_{1}} = 45(-9b_{4}+105b_{6}-280b_{8} .
\end{equation}

For $J=9/2$ (Nd$^{3+}$, Np$^{4+}$, U$^{3+}\ldots$) the multiplet splits into a $\Gamma_{7}$ doublet
and two $\Gamma_{8}$ quartets, respectively splitted by
\begin{equation}
E_{\Gamma_{8}^{(2)}}-E_{\Gamma_{8}^{(1)}} = 15\sqrt{7(103 b_{4}^{2}
+4704 b_{4}b_{6} + 85680 b_{6}^{2}-10584 b_{4}b_{8}+362880b_{6}b_{8}
+3129840b_{8}^{2})}
\end{equation}
and
\begin{equation}
E_{\Gamma_{7}}-E_{\Gamma_{8}^{(1)}} = \frac{735}{2}b_{4}
-12600b_{6}+28350b_{8}+\frac{(E_{\Gamma_{8}^{(2)}}-E_{\Gamma_{8}^{(1)}})}{2}.
\end{equation}

The above results, with their algebraical signs, are correct for
any particular order of the energy levels.

\newpage

\begin{table*}[h]
\caption{\label{tab:table1} $\nu _{k}^{(m,n)}(J)$ coefficients for
values of $J$ corresponding to the ground state of light
lanthanides and actinides.}
\begin{ruledtabular}
\begin{tabular}{cccc}
Coefficient & $J=5/2$ & $J=4$ & $J=9/2$ \\
\hline
$\nu _{4}^{(4,4)}$ & $32/10395$ & $-1328/2760615$ & $-296/920205$ \\
$\nu _{4}^{(6,6)}$ & $64/\left( 33\sqrt{455}\right) $ &
$-\sqrt{2/35}\times
256/184041$ & $-\sqrt{17/7}\times 128/184041$ \\
$\nu _{4}^{(4,6)}$ & $448/6435$ & $-6464/920205$ & $-39488/15643485$ \\
$\nu _{6}^{(4,4)}$ & $0$ & $-1928/93648555$ & $-452/52026975$ \\
$\nu _{6}^{(6,6)}$ & $0$ & $-\sqrt{2/35}\times 32/212355$ & $4/\left( 212355%
\sqrt{119}\right) $ \\
$\nu _{6}^{(4,6)}$ & $0$ & $-512/22297275$ & $-1072/126351225$ \\
$\nu _{8}^{(4,4)}$ & $0$ & $2/36891855$ & $4/184459275$ \\
$\nu _{8}^{(6,6)}$ & $0$ & $-\sqrt{2/35}\times 2/250965$ & $-1/\left( 150579%
\sqrt{119}\right) $ \\
$\nu _{8}^{(4.6)}$ & $0$ & $7/418275$ & $49/14221350$ \\
\end{tabular}
\end{ruledtabular}
\end{table*}

\begin{table}[h]
\caption{\label{tab:table2} Crystal-field parameters found in the
literature for various actinide dioxides. The labels correspond to
those used in Fig.~\ref{fig:fig3}.}
\begin{ruledtabular}
\begin{tabular}{clrrl}
Label & Compound & $V_{4}$ (meV) & $V_{6}$ (meV) & Source \\
\hline
1 & UO$_{2}$ & $-409$ & $24.8$ & Ref.~\onlinecite{rahman} \\
2 & UO$_{2}$ & $-123$ & $26.5$ & Ref.~\onlinecite{amoretti} \\
3 & NpO$_{2}$ & $-104$ & $6.2$ & Ref.~\onlinecite{amorettinpo2} \\
4 & NpO$_{2}$ & $-132$ & $26.4$ & Ref.~\onlinecite{amorettinpo2} \\
5 & PuO$_{2}$ & $-151$ & $31.0$ & Ref.~\onlinecite{kernpuo2} \\
\end{tabular}
\end{ruledtabular}
\end{table}

\newpage

\begin{table}[h]
\caption{\label{tab:table3} The intramultiplet transition energies
experimentally detected by INS for some actinide and rare-earth
dioxides are compared with the corresponding energy splitting
calculated in a $J$-mixing framework with the unique set of
parameters $A_{4}=-19.6~{\rm meV}/a_{0}^{4}$ , $A_{6}=0.666~{\rm
meV}/a_{0}^{6}$ (i.e. the solution labelled 4 in
Fig.~\ref{fig:fig3}).}
\begin{ruledtabular}
\begin{tabular}{lcrr}
Compound & Transition & \multicolumn{2}{c}{Energy (meV)} \\
 & & Exp. (meV) & Calc. (meV) \\
\hline
UO$_{2}$ & $\Gamma _{5}\rightarrow \Gamma _{3}$ & $155$\footnotemark[1] & $167$ \\
UO$_{2}$ & $\Gamma _{5}\rightarrow \Gamma _{4}$ & $172$\footnotemark[1] & $187$ \\
NpO$_{2}$ & $\Gamma _{8}^{(2)}\rightarrow \Gamma _{8}^{(1)}$ & $55$\footnotemark[2] & $56$ \\
PuO$_{2}$ & $\Gamma _{1}\rightarrow \Gamma _{4}$ & $123$\footnotemark[3] & $112$ \\
PrO$_{2}$ & $\Gamma _{8}\rightarrow \Gamma _{7}$ & $131$\footnotemark[4] & $82$ \\
\end{tabular}
\end{ruledtabular}
\footnotetext[1]{Refs. \onlinecite{amoretti} and
\onlinecite{kernuo2}.} \footnotetext[2]{Ref.
\onlinecite{amorettinpo2}.} \footnotetext[3]{Ref.
\onlinecite{kernpuo2}.} \footnotetext[4]{Ref.
\onlinecite{boothroyd}.}
\end{table}

\newpage

\begin{figure}[h]
\caption{\label{fig:fig1} Possible solutions of the equation
$E_{\Gamma_{7}}-E_{\Gamma_{8}} = 131~\rm{meV}$ in terms of the
crystal-field parameters $V_{4}$ and $V_{6}$ of PrO$_{2}$. Full
line: $J$-mixing effects are taken into account by exact
diagonalization of the full $f^{1}$ Hamiltonian
[Eq.~(\ref{f1compl})]. Dashed line: $J$-mixing effects are
neglected. Dashed-dotted line: $J$-mixing effects are taken into
account perturbatively [Eq.~(\ref{f1appr})].}
\caption{\label{fig:fig2} Possible solutions of the disequations:
$165~\rm{meV} \leq E_{\Gamma_{4}}-E_{\Gamma_{5}} \leq
179~\rm{meV}$ and $150~\rm{meV} \leq E_{\Gamma_{3}}-E_{\Gamma_{5}}
\leq 160~\rm{meV}$, in terms of the crystal-field parameters
$A_{4}$ and $A_{6}$ of UO$_{2}$. The dashed ellipse represents the
``safe zone'' defined in the text.}
\caption{\label{fig:fig3} Possible solutions of the equations:
$E_{\Gamma_{8}^{(1)}}-E_{\Gamma_{8}^{(2)}} = 55 \pm 5~\rm{meV}$
(NpO$_{2}$); $E_{\Gamma_{4}}-E_{\Gamma_{1}} = 123 \pm 5~\rm{meV}$
(PuO$_{2}$), in terms of the crystal-field parameters $A_{4}$ and
$A_{6}$. The dashed ellipse near the bottom-right corner
represents the ``safe zone'' for UO$_{2}$ (see
Fig.~\ref{fig:fig2}); the numbered dots are sets of parameters
found in the literature for different compounds (see
Table~\ref{tab:table2} for details).}
\caption{\label{fig:fig4} Experimental (dots,
Refs.~\onlinecite{karraker,moore}) and calculated (lines, present
work) inverse magnetic susceptibility for AmO$_{2}$ and CfO$_{2}$,
with $A_{4}=-19.6~{\rm meV}/a_{0}^{4}$ ; $A_{6}=0.666~{\rm
meV}/a_{0}^{6}$ ; $\chi^{-1}_{{\rm Am}}\left( 0 \right) = 80~{\rm
mol/cm}^{3}$ ; $\chi^{-1}_{{\rm Cf}}\left( 0 \right) = 3~{\rm
mol/cm}^{3}$ .}
\end{figure}


\newpage

\begin{references}

\bibitem{newmanbook} D. J. Newman and B. Ng, {\it Crystal Field Handbook}
(Cambridge University Press, 2000).
\bibitem{liviotti} E. Liviotti, S. Carretta, and G. Amoretti, J. Chem. Phys.
{\bf 117}, 3361 (2002).
\bibitem{prlfe8} S. Carretta, E. Liviotti, N. Magnani, P. Santini
and G. Amoretti, Phys. Rev. Lett. (to be published);
cond-mat/0404011.
\bibitem{magnani} N. Magnani, S. Carretta, E. Liviotti, and G. Amoretti,
Phys. Rev. B {\bf 67}, 144411 (2003).
\bibitem{oct1} P. Santini and G. Amoretti, Phys. Rev. Lett. {\bf
85}, 2188 (2000).
\bibitem{oct2} J. A. Paix\"ao, C. Detlefs, M. J. Longfield, R. Caciuffo, P. Santini, N. Bernhoeft, J. Rebizant,
and G. H. Lander, Phys. Rev. Lett. {\bf 89}, 187202 (2002).
\bibitem{bound} J. M. Fournier, A. Blaise, G. Amoretti, R.
Caciuffo, J. Larroque, M. T. Hutchings, R. Osborn and A. D.
Taylor, Phys. Rev. B {\bf 43}, 1142 (1991).
\bibitem{sasaki} K. Sasaki and Y. Obata, 1970, J. Phys. Soc. Jpn. {\bf 28}, 1157 (1970).
\bibitem{cowley} R. A. Cowley and G. Dolling, Phys. Rev. {\bf 167}, 464
(1968).
\bibitem{uo299} R. Caciuffo, G. Amoretti, P. Santini, G. H.
Lander, J. Kulda and P. de V. Du Plessis, Phys. Rev. B {\bf 59},
13892 (1999).
\bibitem{amoretti} G. Amoretti, A. Blaise, R. Caciuffo, J. M. Fournier, M. T. Hutchings,
R. Osborn, and A. D. Taylor, Phys. Rev. B {\bf 40}, 1856 (1989).
\bibitem{amorettinpo2} G. Amoretti, A. Blaise, R. Caciuffo, D. Di Cola, J. M. Fournier,
M. T. Hutchings, G. H. Lander, R. Osborn, A. Severing, and A. D.
Taylor, J. Phys.: Condens. Matter {\bf 4}, 3459 (1992).
\bibitem{slichter} C. P. Slichter, {\it Principles of Magnetic Resonance}
(Springer-Verlag, Berlin, 1990)
\bibitem{notaopstevens} Stevens' operators of eight order have been defined so that
$\left< J M \left| O_{8}^{q} \right| J M^{\prime} \right> =
\left< J M \left| C_{q}^{(8)} + C_{-q}^{(8)} \right| J M^{\prime} \right>
\times N_{8}^{q} / 2$,
with $N_{8}^{0} = 128$; $N_{8}^{4} = 64\sqrt{2/693}$; $N_{8}^{8} = 128\sqrt{2/6435}$.
The general expression for the matrix elements of $C^{(k)}_{q}$ within
a multiplet of fixed $J$ can be found in: D. Smith and J. H. M. Thornley,
Proc. Phys. Soc. 89, 779 (1966).
\bibitem{boothroyd} A. T. Boothroyd, C. H. Gardiner, S. J. S. Lister, P. Santini,
B. D. Rainford, L. D. Noailles, D. B. Currie, R. S. Eccleston, and R. I. Bewley,
Phys. Rev. Lett. {\bf 86}, 2082 (2001).
\bibitem{wu} X. Wu and A. K. Ray, Eur. Phys. J. B {\bf 19}, 345
(2001); S. Xia and J. C. Krupa, J. Alloys Comp. {\bf 307}, 61
(2000).
\bibitem{raggimedi} J. P. Desclaux and A. J. Freeman, in {\it Handbook of the
Physics and Chemistry of the Actinides}, edited by A. J. Freeman and G. H. Lander
(North-Holland, Amsterdam, 1984), Vol. 1.
\bibitem{kernuo2} S. Kern, C.-K. Loong and G. H. Lander, Phys. Rev. B {\bf 32},
3051 (1985).
\bibitem{rahman} H. U. Rahman and W. A. Runciman, J. Phys. Chem. Solids {\bf 27},
1833 (1966).
\bibitem{raphael} G. Raphael and R. Lallement, Solid State Commun. {\bf 6}, 383 (1968).
\bibitem{kernpuo2} S. Kern, R. A. Robinson, H. Nakotte, G. H. Lander, B. Cort, P. Watson,
and F. A. Vigil, Phys. Rev. B {\bf 59}, 104 (1999).
\bibitem{karraker} D. G. Karraker, J. Chem. Phys. {\bf 63}, 3174 (1975).
\bibitem{moore} J. R. Moore, S. E. Nave, R. G. Haire, and P. G. Huray,
J. Less Common Met. {\bf 121}, 187 (1986).
\bibitem{mamo} M. Colarieti-Tosti, O. Eriksson, L. Nordstr\"om, J. Wills, and M. S. S. Brooks,
Phys. Rev. B {\bf 65}, 195102 (2002).



\end{references}
\end{document}